\documentstyle[12pt]{article}
\begin{document}      
\pagestyle{empty}

\title{ An Analytic Solution of Hydrodynamic Equations with Source Terms in 
        Heavy Ion Collisions }
\author{Pengfei Zhuang$^1$, Zhenwei Yang$^1$, Mei Huang$^2$, Weiqin Zhao$^2$\\
        $^1$ Physics Department, Tsinghua University, Beijing 100084, China\\
        $^2$ Institute of High Energy Physics, Chinese Academy of Sciences,\\
             Beijing 100039,China}
\date{\today}
\maketitle

\begin{abstract}
The energy and baryon densities in heavy ion collisions are estimated by 
analytically solving a 1+1 dimensional hydrodynamical model with source terms. Particularly, 
a competition between the energy and baryon sources and the expansion of the system is discussed in detail.
\end{abstract}

\pagestyle{plain}
\pagenumbering{arabic}
\vspace{0.25cm}

Theoretical evidence from lattice simulations\cite{karsch} of Quantum chromodynamics (QCD) 
for a phase transition from a hadron gas to a quark-gluon plasma at high temperatures has 
prompted experimental efforts to create this new phase of matter in the laboratory during 
the early stages of relativistic heavy-ion collisions. Besides the deconfinement phase 
transition, the medium effects in hot and dense nuclear matter produced in the collisions 
change the characteristics of strong interaction at hadronic level. For instance, the effects lead to the 
transition from the chiral symmetry breaking phase to the phase in which it is 
restored\cite{weise}. Since 1986, oxygen, sulfur, silicon and other beams have been 
successfully accelerated to 14.5 A GeV at the BNL AGS, and to 200 A GeV at the CERN SPS. 
In the past few years heavy-ion collisions are also investigated at the GSI SIS with 
relative low energies of 0.8-1.8 A GeV to study the in-medium effects related to high 
densities. From the above experiments, people have really seen some new phenomena such as 
the $J/\psi$-suppression\cite{gavin}, the dilepton distribution shift\cite{li} and the 
${K^-\over K^+}$ enhancement\cite{gsi}. It is, however, still difficult to say that 
definite signals of formation of quark-gluon plasma in the collisions are found. This 
vague situation is expected to be changed in the near future, since the BNL RHIC will 
start to run soon, providing center of mass energies of up to 100 + 100 A GeV and beams as 
heavy as uranium. To further study the high density effects at hadronic level, a new 
generation of Cooling Storage Ring(CSR) with incident energies of 0.4-0.9 A GeV is 
going to be constructed at LanZhou, China. 

The key point for producing quark-gluon plasma or discussing in-medium effects in 
heavy ion collisions is whether the energy density or baryon 
density achieved in the collisions is high enough. From a general consideration, the energy 
density required for producing the quark-gluon plasma must be several times higher than the 
energy density of normal nuclear matter, and the baryon density for a visible change of 
chiral properties in nuclear matter should be at least the order of the normal nuclear 
density\cite{weise}. Can such a high energy density or baryon density be achieved at SPS 
and RHIC or at AGS, SIS and CSR? The answer to this question depends strongly on the 
degree of nuclear stopping power. The experimental data from AGS\cite{tannenbaum} 
demonstrated quantitatively that around and below the AGS energy any incident nucleus can be 
fully stopped in a heavy target. In contrast with this, the data\cite{heck} from SPS show 
that the incident energy is not fully deposited in the target, the colliding nuclei are 
partially transparent. Since the degree of stopping power is variable with the incident 
energy, it is believed that a baryon-rich matter may be created either in the central region at lower energies or in the fragmentation region at higher energies.  

Assuming full stopping, the maximum energy density and baryon density in the center of 
mass frame for a symmetrical $A-A$ collision can be simply evaluated to be
\begin{eqnarray}
\label{stopping}
&& \epsilon_{max}=2 n_N m_N \cosh^2(y_c)\ ,\nonumber\\
&& n_{max} = 2 n_N \cosh(y_c)\ ,
\end{eqnarray}
where $n_N$ is the density of nucleons in normal nuclei, $m_N$ the mass of the nucleon, 
and $y_c$ the rapidity of the colliding nuclei in the center of mass frame.

For the collisions with partial transparency, Bjorken's formula\cite{bjorken} gives 
\begin{equation}
\label{transparency}
\epsilon_{max}={1\over \pi R_A^2\tau_0}\left({dE_T\over dy}\right)_{max}\ ,
\end{equation}
where $\tau_0$ is the formation time of the secondary particles, 
$R_A$ the radius of the colliding nuclei, and $\left({dE_T\over dy}\right)_{max}$ the 
transverse energy distribution in the central region.

Hydrodynamical models are widely used to describe the macroscopic and collective effects 
in heavy ion collisions, assuming thermal and chemical equilibrium through a 
large number of nucleon-nucleon and secondary collisions. Landau\cite{landau} was the 
first to treat high energy collisions with fluid dynamics, and Bjorken\cite{bjorken} 
considered partial transparency in relativistic collisions. Assuming the colliding system is 
boost invariant, the maximum energy density reached at time $\tau_0$ decreases 
monotonously as\cite{bjorken}
\begin{equation}
\label{nosource1}
\epsilon(\tau) = \epsilon_{max}\left({\tau_0\over \tau}\right)^{1+v_s^2}
\end{equation}
in the central rapidity region, where $v_s$ is the sound velocity in the fluid. 

In Landau and Bjorken's discussions there is no consideration of the initial collision process. Two 
nuclei are assumed to collide at a unique space-time point $(x,t)=(0,0)$. This means that 
all nucleons in the projectile collide with the target at the same time 
$\tau=0$, and the secondary particles from different collisions are formed at the same time $\tau=\tau_0$. Therefore, the maximum energy 
density $\epsilon_{max}$ in (\ref{stopping}) or (\ref{transparency}) is just a simple 
superposition of the $N-A$ collisions. After time $\tau_0$, the energy density drops 
down in terms of (\ref{nosource1}) due to the expansion of the system. 

It is clear that any nucleus-nucleus ($A-A$) collision with a finite rapidity $y_c$ in 
the center of mass frame takes a finite collision time 
\begin{equation}
\label{dt}
d_t={2(R_A-r_N)\over \sinh y_c}\neq 0\ ,
\end{equation}
long or short according to the incident energy and the size of the colliding nuclei. Here $r_N$ is the nucleon radius. Considering a central $^{197}Au-^{197}Au$ collision at incident energies $E_{lab}=0.4-14.5$ A 
GeV, namely from CSR to AGS energies, the collision time $4 fm \leq d_t \leq 23 fm$ is 
comparable with or even larger than the total time of the fluid description of the system. 
However, when we consider the same colliding nuclei but at RHIC energy, the collision time 
is only $0.1 fm$, the neglect of its influence on the fluid evolution is therefore a 
good approximation in high energy limit. In this paper, we will concentrate our 
attention on the dependence of the energy and baryon densities on the collision time for 
heavy ion collisions with full nuclear stopping. 

The boost invariance in the longitudinal direction used to derive the scaling solution (\ref{nosource1}) is a good assumption at early stages of high-energy nuclear collisions when the transverse expansion can be neglected. As the transverse flow develops with increasing time, the fluid cools down more and more rapidly\cite{bjorken}. At the late time in the evolution of the reaction, the fluid will expand radially outward, and one can use boost invariance in the radial direction instead of in the longitudinal direction\cite{lampert}. In fact, the radial scaling solution is widely used in analytic study of relativistic heavy-ion collisions, for instance, the investigation of disoriented chiral condensate\cite{randrup}. Compared with the high-energy collisions, the kinematics dominated longitudinal expansion in low-energy collisions is much weaker, but the transverse expansion does not change a lot. For example, when the colliding energy increases from about $20$ A GeV at SPS to $200$ A GeV at RHIC, the transverse flow velocity changes from $0.55 c$ to $0.65 c$\cite{heinz}. On the other hand, what we are interested in here is the initial collision period. At high energies, this period is very short, the longitudinal scaling solution is a good approximation in the period. At low energies, the collision time $d_t$ has the same order as the total evolution time of the fluid. For most time of the collision period, the transverse flow is comparable with the longitudinal flow. Therefore, for analytically estimating the energy density and baryon number density in low-energy collisions the radial scaling solution seems better than the longitudinal scaling solution. In this case the time evolution of the energy density (\ref{nosource1}) is replaced by
\begin{equation}
\label{nosource2}
\epsilon(\tau) = \epsilon_{max}\left({\tau_0\over\tau}\right)^{3(1+v_s^2)}\ ,
\end{equation}
the factor $3$ in the exponent arises from the 3-dimensional homogeneous expansion.

In principle, the expansion of the system starts at the initial time $\tau_0$. However, in the very early stage the expansion is weaker than the homogeneous evolution (\ref{nosource2}). For the sake of simplicity we still use the homogeneous mechanism as an effective description of the whole expansion process, but the beginning time of the expansion is shifted from $\tau_0$ to $\tau_{exp}$. Since the collision time $d_t$ is large enough, $\tau_{exp}$ can be restricted in the region $\tau_0\le\tau_{exp}\le\tau_0+d_t$.
   
Let us first analyze qualitatively the behavior of the densities when a non-zero collision 
time is taken into account. We still use $N-A$ collisions as input, but the collision time 
of each $N-A$ collision is different. This implies a different time in the 
formation of secondary particles. Those produced by the first $N-A$ collision are formed 
at time $\tau_0$, those by the last one at $\tau_0+d_t$, and others between 
$\tau_0$ and $\tau_0+d_t$. As there is a continuous input of secondaries and no expansion before $\tau_{exp}$,  the energy and baryon densities of the system will increase linearly with time in the range $\tau_0\leq\tau\leq\tau_{exp}$. When the effective expansion begins, there is a competition between the input from the left $N-A$ collisions and the expansion of the system. While the former makes the densities increase, the latter attenuates the fluid strongly. Therefore, the evolution of the densities in the time interval $\tau_{exp}\leq\tau\leq\tau_0+d_t$ depends on the input and expansion rates. After the time $\tau_0+d_t$ when the input from the $N-A$ collisions ends, the densities drop down monotonously like Eq.(\ref{nosource2}).

In the absence of viscosity and thermal conductivity, the behavior of a relativistic fluid 
is controlled by the energy-momentum and baryon number conservation laws\cite{kajantie},
\begin{eqnarray}
\label{source1}
&& \partial_\mu T^{\mu\nu}=\Sigma^\nu\ ,\nonumber\\
&& \partial_\mu n^\mu = \sigma\ ,\ \ \ \ \ \mu,\nu=0,1,2,3
\end{eqnarray}
where $T^{\mu\nu}=-P g^{\mu\nu}+(\epsilon+P)u^\mu u^\nu$ is the energy-momentum tensor, 
$n^\mu = n u^\mu$ the baryon flow, $P=P(\epsilon)$ the pressure, $\epsilon$ the local 
energy density and $u^\mu$ the four-velocity of the fluid. The source terms $\Sigma^\mu$ 
and $\sigma$ indicate the continuous deposition of energy, momentum and baryon number in the 
fluid due to the $N-A$ collisions. In ultra-relativistic collisions, there is almost no net baryon number in the central region, one discusses the energy density only. For collisions with full nuclear stopping, we must consider simultaneously the energy and baryon densities characterized by (\ref{source1}).

Using two Lorentz covariant coordinate, the proper time 
$\tau = \sqrt{t^2-r^2}$ and the space-time rapidity $y={1\over 2}\ln{t+r\over t-r}$ 
instead of $r$ and $t$, the above hydrodynamical equations are greatly simplified as
\begin{eqnarray}
\label{source2}
\partial_\tau\epsilon + 3{P+\epsilon\over \tau} = \Sigma,\nonumber\\
\partial_\tau n + 3{n\over\tau}=\sigma\ ,
\end{eqnarray}
where the source terms on the right side mean the energy density and baryon density 
deposited in the fluid per unit time. Since the total energy input and the total baryon number input during the collision time $d_t$ are $\epsilon_{max}(A^{1/3}-1)/A^{1/3}$ and 
$n_{max}(A^{1/3}-1)/A^{1/3}$, where $\epsilon_{max}$ and $n_{max}$ are shown in (\ref{stopping}), the source terms are given by
\begin{eqnarray}
\label{source3}
&& \Sigma = {\epsilon_{max}(A^{1/3}-1)\over A^{1/3} d_t}\ ,\nonumber\\
&& \sigma = {n_{max}(A^{1/3}-1)\over A^{1/3} d_t}\ .
\end{eqnarray}
Taking into account the initial condition
\begin{eqnarray}
\label{source4}
&& \epsilon(\tau_0) = {\epsilon_{max}\over A^{1/3}}\ ,\nonumber\\
&& n(\tau_0) = {n_{max}\over A^{1/3}}\ ,
\end{eqnarray}
the solution of the differential equations (\ref{source2}) is
\begin{eqnarray}
\label{source5}
&& \epsilon(\tau)=\left\{\begin{array}{ll}
                         \epsilon(\tau_0)+(\tau-\tau_0)\Sigma &\ \ \ \ \ 
                         \tau_0\leq\tau\leq\tau_{exp}\ ,\\
                         \epsilon(\tau_{exp})\left({\tau_{exp}\over \tau}\right)^{3(1+v_s^2)}+
                         {\Sigma\over 3(1+v_s^2)+1}\tau
                         \left(1-\left({\tau_{exp}\over\tau}\right)^{3(1+v_s^2)+1}\right) 
                         &\ \ \ \ \ \tau_{exp}\leq\tau\leq\tau_0+d_t\ ,\\
                         \epsilon(\tau_0+d_t)\left({\tau_0+d_t\over\tau}\right)^{3(1+v_s^2)} 
                         &\ \ \ \ \ \ \tau_0+d_t\leq\tau \ ,
                         \end{array}\right.\nonumber\\
&& \nonumber\\
&& n(\tau)=\left\{\begin{array}{ll}
                         n(\tau_0)+(\tau-\tau_0)\sigma 
                         &\ \ \ \ \ \tau_0\leq\tau\leq\tau_{exp}\ ,\\
                         n(\tau_{exp})\left({\tau_{exp}\over \tau}\right)^3+{\sigma\over                                 4}\tau\left(1-\left({\tau_{exp}\over\tau}\right)^4\right)
                         &\ \ \ \ \ \tau_{exp}\leq\tau\leq\tau_0+d_t\ ,\\
                         n(\tau_0+d_t)\left({\tau_0+d_t\over\tau}\right)^3 
                         & \ \ \ \ \ \tau_0+d_t\leq\tau\ ,
                         \end{array}\right.
\end{eqnarray}
where the sound velocity $v_s$ comes from the equation of state $P=v_s^2\epsilon$. Before the effective expansion begins, the energy density increases linearly. In the 
time period $\tau_{exp}\leq\tau\leq\tau_0+d_t$, the energy density contains two terms. The 
first one is the time evolution of the density at $\tau_{exp}$, and the second term reflects the 
contribution of the source term. After $\tau_0+d_t$ there is no more energy input, the second term disappears and the 
density decreases with time. The behavior of the baryon density is similar to that of the 
energy density. Obviously the heavier the colliding nuclei are or the lower the incident 
energy is, the longer is the collision time $d_t$ and the more important is the energy and 
baryon diffusion. 

Let us now turn to numerical results. As we have emphasized above, the effect of finite 
collision time on the density evolution is especially important for collisions with 
full stopping. In our numerical calculations below we take 
$^{197}Au-^{197}Au$ collisions at AGS, SIS and CSR energies, namely from $0.4$ to $14.5$ A 
GeV. At these low energies, while it seems not possible to create quark-gluon plasma in 
the collisions, the in-medium effects at hadronic level are still visible if the energy 
density or baryon density is larger than the relevant density of normal nuclear matter. 
For instance, from a model independent calculation\cite{weise} the broken chiral 
symmetry in the vacuum $(T=n=0)$ is partially restored even in the normal nuclear matter. 
The energy density and baryon density that are shown in the following are scaled, 
respectively, by the normal energy density $\epsilon_N=m_N n_N$ and normal baryon density 
$n_N=1/\left({4\over 3}\pi r_N^3\right)$, where the nucleon mass $m_N$ and the nucleon 
radius $r_N$ are taken to be $0.94$ GeV and $1.1$ fm. For a perfect fluid, the equation of 
state can be written as $P={1\over 3}\epsilon$. In real fluid with finite hadron mass, 
viscosity and thermal conductivity, one has $P=v_s^2\epsilon$ with $v_s^2<{1\over 3}$. In 
our calculation we take $v_s^2={1\over 5}$.

In Fig.1a, the energy density, scaled by $\epsilon_N$, is shown as a function of the proper time $\tau$ for four values of the expansion time $\tau_{exp}$ at the incident energy $E_{lab}=1.8$ A GeV. For $\tau_{exp} = \tau_0$, the effective expansion and the energy input start simultaneously. The energy density drops down in the very beginning when the energy input rate is less than the expansion rate, and then goes up smoothly when the energy deposited in the fluid is high enough. The maximum energy density $\epsilon(\tau_0+d_t)/\epsilon_N = 0.76$ is larger than the initial value $\epsilon(\tau_0)/\epsilon_N = 0.67$. When the fluid evolves beyond the source region, the energy density decreases monotonously. As the effective expansion starts later than the formation of the fluid, $\tau_{exp}>\tau_0$, the energy density increases linearly before the expansion. The increase or decrease of the energy density in the region $\tau_{exp}\leq\tau\leq\tau_0+d_t$ depends on the competition between the amount of the energy deposited in the fluid and the expansion rate. When $\tau_{exp}$ is larger than $1.3\tau_0$, the position of the maximum energy density moves from $\tau_0+d_t$ to $\tau_{exp}$. For $\tau_{exp}\geq \tau_0+d_t$, there is no expansion in the source region, the maximum energy density $\epsilon(\tau_0+d_t)$ achieves its limit value (\ref{stopping}).

The time evolution of the baryon density, scaled by $n_N$, is shown in Fig.1b. The structure is very similar to that of the energy density.

In Fig.2, we show the maximum energy density and baryon density, scaled by $\epsilon_N$ 
and $n_N$, as functions of incident energy $E_{lab}$ for the four values of the expansion time $\tau_{exp}$. In the case of no expansion in the source region , the maximum energy density and baryon density given by (\ref{stopping}) increase very rapidly and reach $17.5$ and $6$ at $E_{lab}=14.5$ A GeV. When 
the expansion in the source region is included, the maximum densities are strongly suppressed by the expansion.

Obviously the energy and baryon densities in heavy ion collisions can not reach the values estimated in the case of no expansion in the source region. The attenuation of the fluid in the initial collision period must be considered. As is well known, the in-medium effects in low energy collisions are mainly related to the baryon number density. When the in-medium effects become visible, the baryon density should be at least larger than the normal nuclear density, $n_{max}/n_N\geq 1$. The experimental data of heavy ion collisions at SIS with incident energies $E_{lab}=0.8-1.8$ A GeV, such as the $K^-/K^+$ enhancement\cite{gsi}, do show the importance of the in-medium effects. This means that the baryon density at SIS energies is already beyond the normal nuclear density. From the comparison of this analysis with our calculations shown in Figs.1 and 2, we can choose approximately $\tau_0+d_t/2$ as the beginning time of the effective expansion for the estimation of the energy and baryon densities in low energy collisions. 

In summary, we have investigated the influence of non-zero collision time on the evolution 
of energy and baryon densities in heavy ion collisions. When the size of the 
colliding nuclei is large enough and the incident energy is not very high, for instance 
the collisions at AGS, SIS and CSR, the initial collision time is long, and the colliding energy is deposited in the fluid gradually instead of an instantaneous input in ultra-relativistic collisions at RHIC and LHC. In the collision period, there exists a competition between the energy and baryon deposition and the expansion. The densities inside and outside the source region behavior very differently. They are strongly suppressed by the expansion in the source region, and the time to reach the maxima is also remarkably delayed. 
  
\vspace{0.25cm}
The work was supported in part by the NSFC under grant numbers 19845001 and 19925519, and by the Major State Basic Research Development Program under contract number G2000077407.

\begin{center}
{\bf FIGURE CAPTIONS}
\end{center}
Fig.1\\
The time evolution of the energy density (Fig.1a) and baryon density (Fig.1b), scaled by 
$\epsilon_N$ and $n_N$, in $^{197}Au-^{197}Au$ collisions at incident energy $E_{lab}=1.8$ A GeV for four values of the expansion time $\tau_{exp}$. \\

\noindent Fig.2\\
The maximum energy density (Fig.2a) and baryon density (Fig.2b), scaled by $\epsilon_N$ and $n_N$, as functions of incident energy $E_{lab}$ for four values of the expansion time $\tau_{exp}$. 

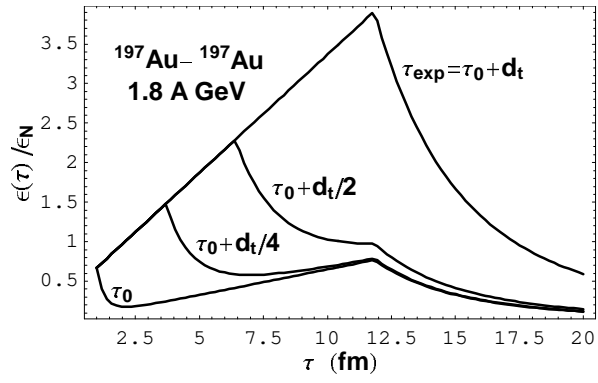
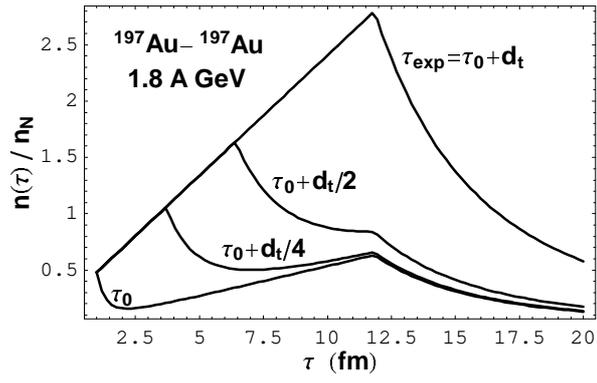
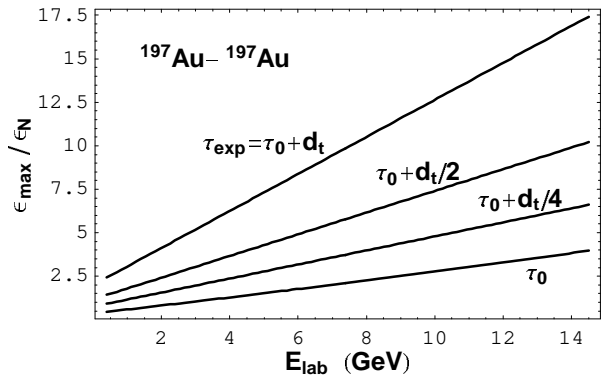
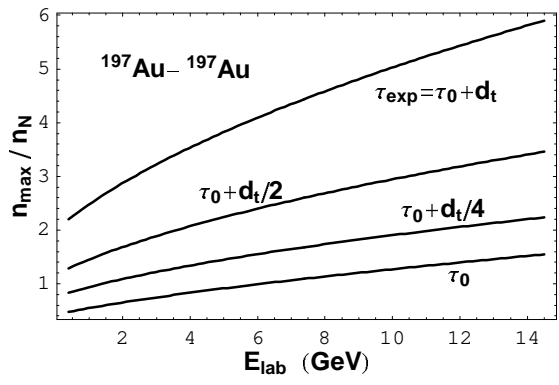
\end{document}